# TARGET MARKET RISK EVALUATION

Anda GHEORGHIU, Associate Professor, Anca GHEORGHIU, Professor and

Ion SPÂNULESCU, Professor, President of Hyperion University

Universitatea Hyperion din București
Calea Călărașilor 169, sector 3, București

## ABSTRACT

*After the shocking series of bankruptcies started in 2008, the public does not trust anymore the classical methods of assessing business risks. The global economic severe downturn caused demand for both developed and emerging economies' exports to drop and the crisis became truly global.*
*However, this current crisis offers opportunities for those companies able to play well their cards. Entering new markets has always been a hazardous entrepreneurial attempt, but also a rewarding one, in the case of success.*
*The paper presents a new indicator meant for assessing the prospective of success or failure for a company trying to enter a new market by using an associative strategy.*
*In order to take the right decision concerning the optimal market entry strategy, marketers may use a software application, "AnBilanț", created by a research team from Hyperion University.*

## Key words

Risk, market entry, investment, evaluation, Discounted Cash Flow, Adjusted Net Asset

## Introduction

People seem to fail to remember about the probability of crisis outbreak in times of prosperity and fail to take quick crisis management actions. Coming out of the current economic crisis, the world has a historical chance to reshape its policies, architecture and institutions and support balanced global growth and financial stability. Especially the emerging economies should avoid macroeconomic imbalances and follow a sustainable growth trend backed by structural reforms. The current financial crisis has highlighted the need for up-to-date and transparent information by type of instrument, currency, creditors, and debtors.

Nevertheless, the crisis offers opportunities for those companies which aim to enter new markets, or to buy cheaply assets and companies in distress. Purchasing at a bargain price has always been a risky attempt, but in most cases a rewarding one, if the return of investment is satisfactory.

**I. An overview over the risks or potential chances to develop a business at a global scale**

The firm internationalization process is habitually accomplished in a gradual manner, by achieving several stages in accord with the motivation to enter and develop globally. There are four stages of internationalization: the exporter, the international firm, the multinational firm and finally, the transnational firm.

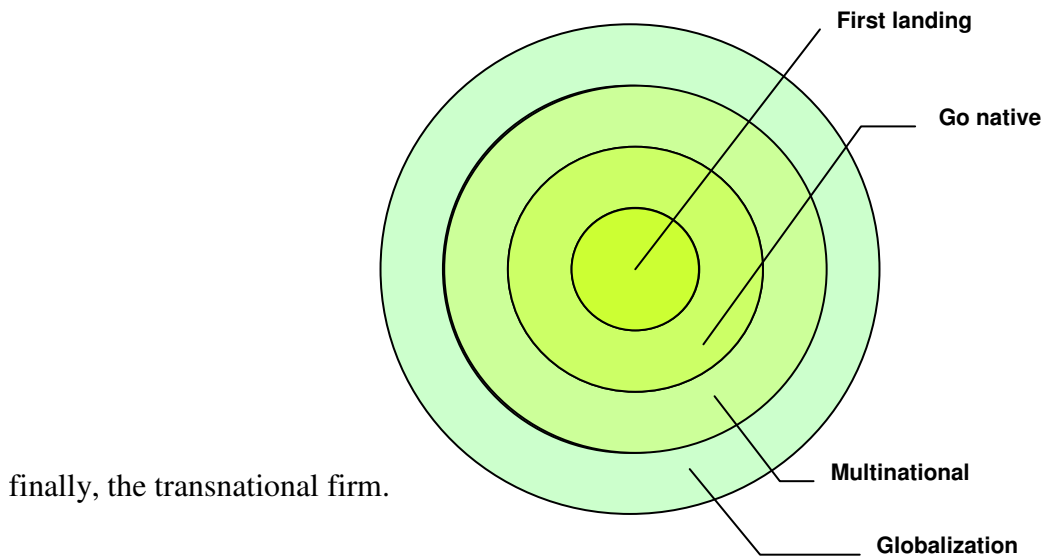

**Figure 1. Stages of the Internationalization Process**

Source: the Authors

Each stage corresponds to the similar four phases of the internationalization process:
- the early internationalization phase,
- production cooperation/local implementation,
- multi-nationalization and
- globalization.

Getting into a new market is, of course, a difficult process, because of the numerous barriers of the internationalization, all being related to the commercial field, to the competitive filed, to the costs resulted from the changing of the suppliers and the products, and to the government policy. There might also be technical barriers, or barriers created by the geographical distance, by the business infrastructure, by the local customs, by the prohibitions, by the customs restrictions, by the specific requests related to the capital. Other aspects that make difficult the process of getting into a foreign market are the languages, the culture, the legislative instability or the lack of the laws for these sensitive domains, or the corruption phenomenon.

Often, the firms start their international process through export operations, which frequently sustain an important weight of their activities. The risks associated with the export are: the risks related to the physical integrity of the wares, the juridical risks, the exchange rate risks, the currency transfer risks, the risks of default, etc.

The associations with other firms for producing and commercialization of goods or services usually have better perspectives to obtain benefits than in the case of exports, but there are a series of disadvantages which could not be neglected, such as: a diminished control of the partners' activities, the possibility that, in time, the partners may become competitors, the misunderstandings concerning the policy of investments, the marketing, etc. These strategies take different forms: licensing, franchising, management contracts, strategic alliances, complex dynamic networks.

The foreign direct investment (FDI) is defined as the establishment of an enterprise by a foreign person or company. The FDI relationship consists of a parent enterprise and a foreign affiliate which together form an international business or a multinational corporation. In order to qualify as FDI the investment must afford the parent enterprise *control* over its foreign affiliate, i.e. owning 10% or more of the ordinary shares or voting power of an incorporated firm or its equivalent for an unincorporated firm; lower ownership shares are considered as portfolio investment.

In the case of a FDI, the firm has the possibility to get the entire control over the operations, it may develop production and marketing policies on long term, it may also obtain reduced production costs by using the local work-force and the raw materials from proximity, it may become the beneficiary of the incentives given by the host-state, and also of the positive image because of the new jobs created by the firm. Even so, the risks related to direct investments are big enough: the currency restrictions, the depreciation of the national currency or, the worst situation, the expropriation. The country risk surveillance is fundamental for taking the best investment decision or for production delocalization in a certain country or region.

Country risk refers to the likelihood that changes in the business environment unfavorably affect operating profits or the value of assets in a specific country. Country risk is a complex term, which includes political risks, financial factors such as currency controls, devaluation or regulatory changes and social factors such as riots or civil war.

Today, managers have a lot of tools to measure the magnitude of risks, among them the most famous are those provided by credit rating agencies, which use quantitative econometric models and focus on financial analysis, whereas political risk providers tend to use qualitative methods, focusing on political analysis. For instance, the rating agencies Moody's and Standard & Poor's make a hierarchy of the countries for their ability to fulfill financial debts and for the bonds quality. The rating's assessment is standardized, by using letters for each risk class, like in the table 1.

From the rating agencies prospective, investors should be aware of threats or opportunities of different markets. Romania's country rating increased after 2001, as Moody's reports state. In 2006, Moody's upgraded the country risk level for Romania (Baa3) for long term credits, with only one rank below the grade considered safe by investors (investment grade). For the other categories, the prospects were also positive, Ba2 for bank deposits in hard currency, Ba1 for long term local debts.

Despite the worst economic crisis after WWII, Moody's agency has affirmed in 2009 the Romania at Baa3; outlook stable. [1] However, Kenneth Orchard, analyst Moody's Investor

---

[1] http://www.moodys.com/

Service, predicted at the end of 2008 in an interview for Money Channel TV an economic recession for Romania in 2009 with a GDP decrease of 0.3% (a mild forecast…).[2]

The agency took into account the global lack of liquidity and the decrease in growth of the main European economies.

| Moody's | Bonds Quality | S&P | Ability to fulfill financial debts |
|---|---|---|---|
| Aaa | Best quality | AAA | Extremely strong |
| Aa | High quality | AA | Very strong |
| A | High average quality Citate | A | Strong – but sensitive on economic |
| Baa | Average quality | BBB | Adequate - but sensitive on |
| Ba | Speculative quality | BB | Less vulnerable - but uncertain in |
| B | Without investment characteristics caracteristici | B | Vulnerable – but for the moment respects the terms of commitment |
| Caa | Low level quality | CCC | Vulnerable |
| Ca | Speculative quality | CC | Very vulnerable at present |
| C | Lowest quality | C | Payment difficulties – but for the moment, payments continue |
|   |   | D | Major payments defaults |

**Table 1.  Grades on Standard and Poor's şi Moody's scale**

Source: Scott David, Understanding and managing investment risk & return, 1990, pag.188

| Date | Rating for long term credits | Trend | Rating for bank deposits in foreign currencies | Trend |
|---|---|---|---|---|
| 06.03.1996 | Ba3 |  | - |  |
| 23.12.1996 | Ba3 | stationary | B1 |  |
| 14.09.1998 | B1 | ▼ | B2 | ▼ |
| 06.11.1998 | B3 | ▼ | Caa1 | ▼ |
| 19.12.2001 | B2 | ▲ | B3 | ▲ |
| 16.12.2002 | B1 | ▲ | B2 | ▲ |
| 11.12.2003 | Ba3 | ▲ | B1 | ▲ |
| 06.10.2006 | Baa3 | ▲ | B1 | stationary |
| 20.03.2009 | Baa3 | ▲ | B1 | stationary |

**Table 2. Evolution of country risk for Romania according to Moody's Investors Service**
Source: Moody's Investors service

---

[2] http://english.hotnews.ro/stiri-business-5164715-moodys-predicts-recession-for-romania-2009.htm

## II. A new tool meant for assessing the prospective of success or failure for a company trying to enter a new market by using an associative strategy

Under such a hostile environment, it becomes more and more necessary to use sophisticated tools for forecasting the risks or potential chances to develop a business at a global scale. The endeavor to enter new markets has always been a risky entrepreneurial act, but also a rewarding one, in the case of success. Producing profits from selling or producing abroad are great accomplishments for firms trying to expand their market share in foreign countries. Nevertheless, companies should think very carefully their market entry strategies, in order to circumvent the negative impact of some risks like: country risk, contractual risks, currency risk, environmental risks, etc.

An useful instrument for assessing the prospective of success or failure for a company trying to enter a new market by using an associative strategy is I, a risk indicator defined[3] as follows:

$$I = N \times F \times \frac{1+RI_T}{1+RI_O} \times \frac{1+RCE_T}{1+RCE_O} \times \frac{V_{\text{company X}}}{CS_{\text{company X}}}$$

where the factors are:
- -the country-risk of the target-market rating (N)
- -the degree of cultural and organizational compatibility (F)
- -the inflation rate for the target-market ($RI_T$)
- -the inflation rate for the country of origin ($RI_O$)
- -the rate of economic growth for the target-country ($RCE_T$)
- -the rate of economic growth for the country of origin ($RCE_O$)
- -the social capital of the X firm, the patrimonial entity that is used for settling an associative strategy ($CS_{\text{X firm}}$)
- -the economic value of the enterprise ($V_{\text{X firm}}$).

If the society is listed at the stock exchange, than $V_{\text{X firm}}$ is the market value.

The rating N assesses the risk degree of the target-country; it can take values between 1-10 (10 for stabile countries and with a good quality of the credit and 1 for the economic crisis situation and incertitude concerning payments and disbursements).

F is a more complex factor and can have values between 0.1 (absolutely incompatible) and 100 (total compatibility).

The ratio $\frac{1+RI_T}{1+RI_O}$ reflects the monetary risks; if $RI_O > RI_T$, it is a signal that the country of origin is weaker than that of the target-market. The reasonable limits of the scale are 0 and 2; over this value, the inflation of the target-market is soaring, the market entry risk being very high.

The ratio $\frac{1+RCE_T}{1+RCE_O}$ reflects the risk of entering a market with a different rhythm of growth than that of the country from where the investment comes. The reasonable limits of the scale are 0 and 2; over this value, the inflation of the target-market is plummeting, the market-entry risk being too high.

The ratio $\frac{V_{\text{company x}}}{CS_{\text{company x}}}$, varies between 0 şi 100.

---

[3] Gheorghiu Anda, *Managementul riscului la pătrunderea pe piața internațională*, Editura Victor, București, 2009, pag.251

If $V_{company\ X} > CS_{company\ X}$, the company is rich in assets, which exceed the scriptural value of the capital (most common situation), while if $V_{company\ X} < CS_{company\ X}$, the company is almost bankrupt.

The indicator I can take positive values (from 0 to $+\infty$).

For a nuanced analysis, one can apply apply the logarithm over I and the result is

$$I^* = \lg I = \lg (N \times \frac{1+RI_T}{1+RI_O} \times \frac{1+RCE_T}{1+RCE_O}) + \lg (F \times \frac{V\ company\ x}{CS\ company\ x}).$$

The first term characterizes the degree of risk of the target region/country, while the second characterizes the microeconomic risk.

If $I^* < 0$ and the country risk is more than 6, the factors which characterize the external environment being in normal limits, than the company envisaged for association is either less evaluated, or almost bankrupt, vulnerable, able to be taken over very easily and changed radically, in terms of items produced.

If $I^* > 5$, under the same circumstances, than the company envisaged for association has a very good financial situation.

Therefore, five intervals of values for the $I^*$ indicator have been settled. The extremes of this grid are:

$I^* < 0$, in this case, the optimal strategy is the direct greenfield investment

$I^* > 5$, in this case, the optimal strategy is the export, as it can be seen in the following table:

| The values of $I^* = \lg(I)$ | The evaluation of the microeconomic environment | The optimal entry strategy |
|---|---|---|
| $I^* < 0$ | The microeconomic environment likely to be entirely taken over | Direct greenfield investment |
| $0 < I^* < 1,6$ | The microeconomic environment likely to be entirely taken over by a buy of the majority of stocks and joining the management team | Acquisition |
| $1,6 < I^* < 2$ | The microeconomic environment likely to be taken over at a equal rate to that of the partner | Mergers, acquisitions |
| $2 < I^* < 5$ | The microeconomic environment favorable for economic cooperation | Licensing, franchising, strategic alliances, management contract |
| $I^* > 5$ | The microeconomic environment hard to be approached through a partnership but favorable for trading operations | Export |

**Table 3. The evaluation of the microeconomic environment analyzed in rapport with the values on the grid of $I^*$**

Source: Gheorghiu Anda, PhD thesis, 2008
http://www.biblioteca.ase.ro/resurse/resurse_electronice

### III. Methods of assessing companies which are not listed at the stock exchange

How should one proceed when the targeted company is not listed at the stock exchange and its current real value does result neither from its stock exchange capitalization, nor from the balance sheet? In this case the targeted company should be approximated through methods specific to the valuation as the Adjusted Net Assets Method or Discounted Cash-Flow.

The Adjusted Net Assets Method (abbreviated here as ***ANC***) requires the following steps:
a) getting the balance sheet reflecting the patrimony at the date of valuation
b) valuating the assets and debts, meaning:
- either to consider the accounting value as the true and fair one, without making any correction over it;
- or converting the accounting value of assets and liabilities at the right market value;

c) recording at the date of valuation the assets and liabilities not shown in the balance sheet;
d) computing the Adjusted Net Assets value.

The *ANC* method requires a great amount of work, because it supposes the splitting up of a company value into distinct parts and the assessment for each of them. Hence, the value of asset-based analysis of a business is equal to the sum of its parts.

The asset valuation approach is based on the principle of substitution: no rational investor will pay more for the business assets than the cost of procuring assets of similar economic utility.

Pursuant to the Generally Accepted Accounting Principles states than an asset or liability should be initially recorded and reported at its original (historical) cost; the theory is that historical costs are easier to verify than are current values, i.e., a market value can only be proven when a sale is consummated). [4] Accordingly, companies have to book and report based on acquisition costs rather than fair market value for most assets and liabilities, most assets are reported in the books at their acquisition value, minus the depreciation, where applicable. However, these values must be adjusted to fair market value wherever possible.

The Discounted Cash-Flow Method (abbreviated here as ***DCF***) is an income-based approach and a way to estimate the value of a company by using the concepts of the time value of money. All future cash flows are estimated and discounted to give their present values. The discount rate used is generally the appropriate cost of capital and may incorporate judgments of the uncertainty (of the future cash flows.

The discounted present value may be expressed as:

$$V_{DCF} = \frac{FV}{(1+i)^n} = FV(1-r)^n$$

where
- $V_{DCF}$ is the discounted present value of the future cash flow (FV);
- FV is the nominal value of a cash flow amount in a future period;
- *i* is the interest rate, which reflects the cost of tying up capital and may also allow for the risk that the payment may not be received in full;
- *r* is the discount rate, which is $i/(1+i)$, i.e. the interest rate expressed as a deduction at the beginning of the year instead of an addition at the end of the year;
- *n* is the time in years before the future cash flow occurs.

We have selected as a case study a company which is listed at the Romanian Stock Exchange (BVB-Bursa de Valori București), Compa Sibiu SA, a producer of spare parts for the automotive industry and in the case of certain products it is the only domestic producer. The users of Compa's products are both domestic producers of cars, commercial vehicles, trucks, tractors, farm implements and railroad rolling stock, firms from the spare part market as well as a range of external clients from countries such as Germany, France, Italy, Hungary, Yugoslavia and U.S.A. Its clients are famous brands such Dacia Renault, Honeywell, Bosch, Piroux, Daikin, Delphi. The company has two major clients, Honeywell and Bosch, representing 63% of its sales in 2007.

---
[4] Griffin P. Michael, MBA Fundamentals: accounting and finance, Kaplan Publishing, New York, S.U.A., 2009, pag.13

Although it is not really "orthodox" to assess a company, which is already quoted over the capital market, we have chosen this approach in order to verify if the hypotheses of the proposed method are true. A company is considered as fairly evaluated over a mature market and the above-mentioned methods (*ANC* and *DCF)* should, theoretically, present as outcome similar values.

In November 2007, the company increased its capital from 14,86 million lei to 21,88 million lei (a 6,46 million euro increase). The main shareholder was the Employees Association, holding 54,59% of the shares.

The Adjusted Net Assets value resulted from the balance sheet as of December 31, 2007, is **361.656.741 lei**. In the *ANC* method version, the indicator I becomes:

$$I = N \times F \times \frac{1+RI_T}{1+RI_O} \times \frac{1+RCE_T}{1+RCE_O} \times \frac{V_{\text{company X}}}{CS_{\text{company X}}} = 7 \times 1 \times \frac{1+0,1}{1+0,04} \times \frac{1+0,05}{1+0,01} \times \frac{361.656.741 \text{ lei}}{21.882.104 \text{ lei}} =$$
$= 7 \times 1 \times 1,06 \times 1,04 \times 16,53 = 127.5587$
$I^* = \lg(I) = 2.105710084531$

| | |
|---|---|
| **I =** | 127.5587 |
| **I\*=lg(I)** | 2.10571008 |

In the case of *DCF* method, the discount rate taken into account was 5% for a 5 years time and with a residual value calculated according to the formula of Gordon-Shapiro[5]:

$$V_{DCF} = \sum_{i=1}^{n} \frac{CFNI_i}{(1+r)^i} + \frac{V_{rez}}{(1+r)^n}$$

where:

$CFNI_n$ is the net cash-flow-ul net from the year *n* (its value was taken from the Profit and Loss account at December 31, 2007) as a difference between revenues and expenses, meaning 7,570,903 lei);

*g* – the rate of perpetual increase of dividends, cautiously considered as 1%;
*r* – the discount rate, 5%.

The residual value is, therefore, $V_{rez} = \frac{CFNI_{n+1}}{r-g} = \frac{CFNI_n(1+g)}{r-g}$.

The value of the company, calculated according to the *DCF* method, was **207.360.284 lei**.

$$\frac{CFNI_i}{(1+r)^i} = \frac{\text{Net profit}_{\text{reference year}}(1+g)^i}{(1+r)^i}, i = 1,2,...n$$

Of course, the analyst has to choose the most appropriate value and in this case, *ANC method* was considered as the accurate one, meaning *361,656,741 lei*. This choice was based on the fact that the use of DCF-type methods in an emerging economy involves many uncertainties, many risks, because the economic environment is quite unstable.

For
-an update rate of r = 5%
-a perpetual dividend growth rate g = 1%
-the country's rating = 7
-the cultural and organizational compatibility score = 1
-a forecasted growth rate of 5% for Romania

---

[5] Stan Sorin, Evaluarea întreprinderii, Editura IROVAL, București, 2003, pag.286

-a growth rate of 1% for EU
-inflation for Romania: 10%
-inflation for EU: 4%

| | |
|---|---|
| **I =** | 127,2133818 |
| **I*=lg(I)** | 2.104532798 |

For comparison, by processing the stock exchange data for the same company, the following results have been obtained:

| | |
|---|---|
| **I =** | 92,36481 |
| **I*=lg(I)** | 1,965507 |

So, the value of I * is:
- 1.965507 by processing the stock exchange data
- 2.10571008 by using the Adjusted Net Assets Method (ANC)
- 2.104532798 by using the capitalization method -Discounted Cash-Flow (DCF).
So, the difference between the two methods (processing the stock exchange data and ANC) is 0.139026.
By considering DCF method, the difference is minor (0.102545 in the favor of processing the stock exchange method). Therefore, the indicator is a powerful and reliable instrument for companies motivated to enter new markets.

## IV. The AnBilanț software

In order to take the right decision concerning the international market entry and to choose the optimal strategy, marketers may use a software application, "AnBilanț", created by a research team from Hyperion University. It is realized under Visual Basic 6 as an executable and it runs under Windows (98/2000/XP) operating system.

Visual Basic[6] is a programming language designed to create applications that work with Microsoft framework; since it provides advanced techniques for visual programming, Visual Basic eases the quick writing of software programs. It is also able to recognize and interact with various types of data files or Database Management Systems (i.e. software systems that allows users to save retrieve and modify information) such as Microsoft Access, dBase, FoxPro, Visual FoxPro, Paradox, SQL Server, etc; it allows the access to documents and Internet/Intranet applications.[7]

AnBilanț eases the decision process in the case of unlisted stock companies, which should be evaluated by using one of the assessment methods presented beforehand.
The application is structured in four areas namely:
I. *The input area of comparative parameters between the target market and the country of origin, to be exact:*
- N-the country-risk of the target-market rating with values between 1-10 (10 for financially solid countries
- F- the degree of cultural and organizational compatibility (F), with values between 0.1 (incompatibility) and 100 (full compatibility)

---

[6] Patrick Tim, Roman Steven, Petrusha Ron, Lomax Paul, Visual Basic 2005 in a nutshell, a desktop quick reference, 3-rd Edition, Ed. O'Reilly, 2006, pag.8
[7] Gheorghiu Anca, Programarea calculatoarelor electronice, Editura Victor, București, 2003, pag.274

- $RI_T$ - the inflation rate for the target-market
- $RI_O$ - the inflation rate for the country of origin
- $RCE_T$ - the rate of economic growth for the target-country
- $RCE_O$ - the rate of economic growth for the country of origin
- r -the discount rate, i.e. the rate at which costs and estimated future incomes of the investment are discounted to calculate the present value of it
- g- dividend perpetual growth rate.

II. *The area that displays the values calculated according to the Discounted Cash Flow method (DCF) and Adjusted Net Asset method (ANC).* After selecting the amount considered by the assessor as the most appropriate the software automatically calculates the indices I and I *= log (I), which asses the market environment of the target country / region and deliver the strategic investment recommendation.

III. *The area of financial analysis* applied to items selected from the balance sheet or profit and loss account, accompanied by graphic illustration of the dynamics of the economic factors.

IV. *Copyright.*

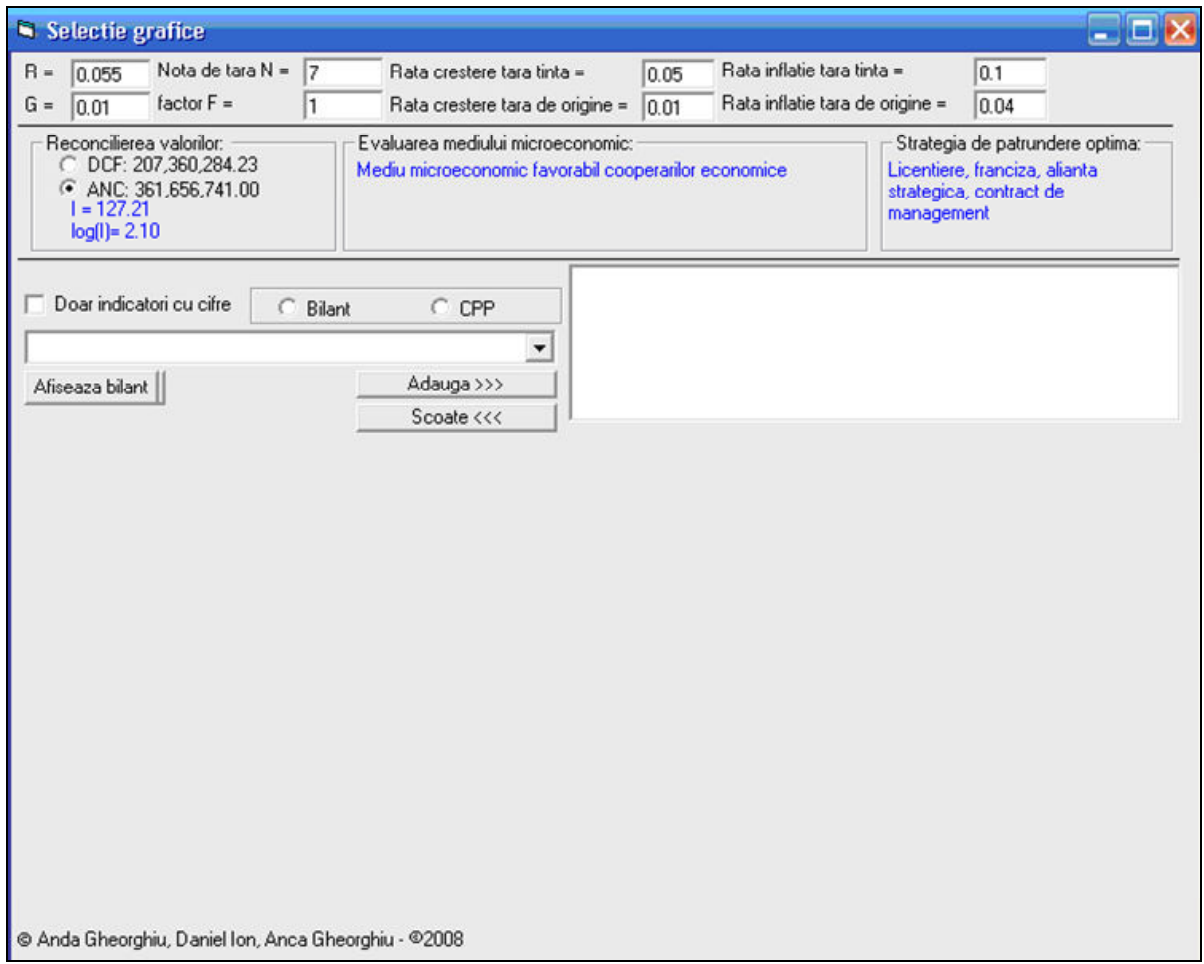

**Figure 2. The visual appearance of the application AnBilanț**

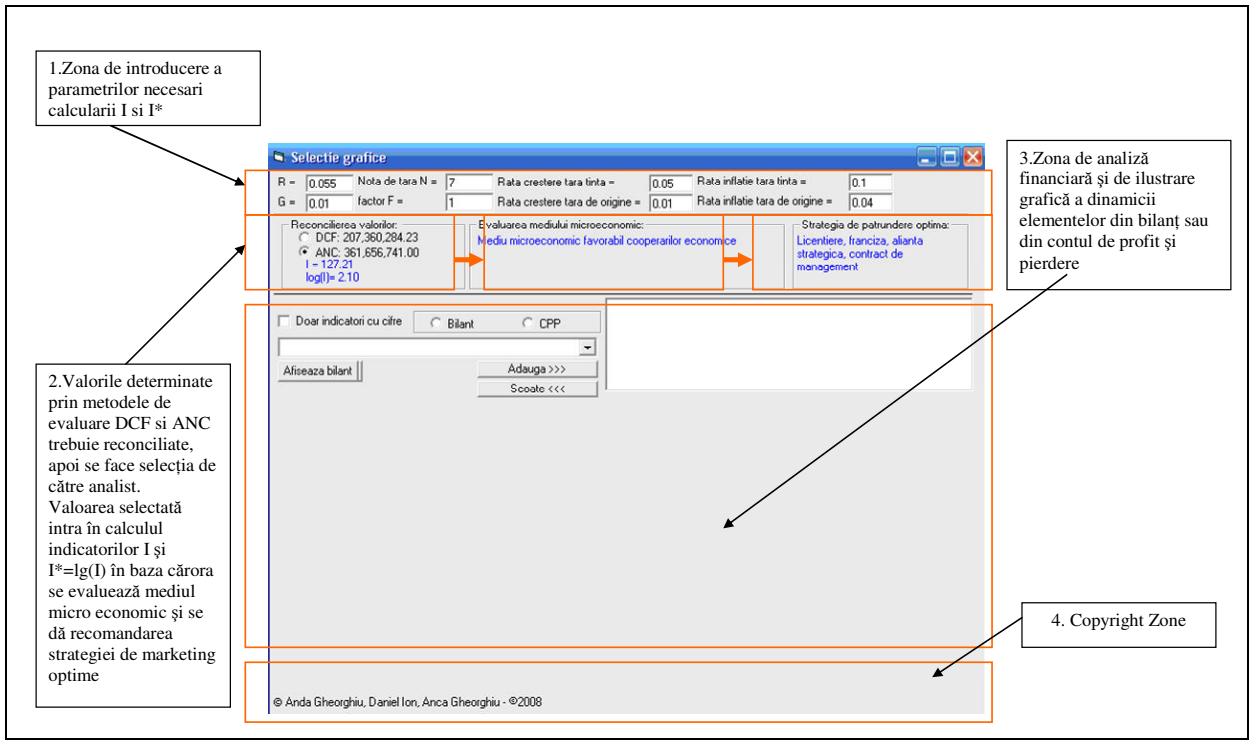

**Figure 3. The structure of menu in the application AnBilanț**

The rates (r and g) and the factors which characterize the target/origin market are illustrated in the figure below:

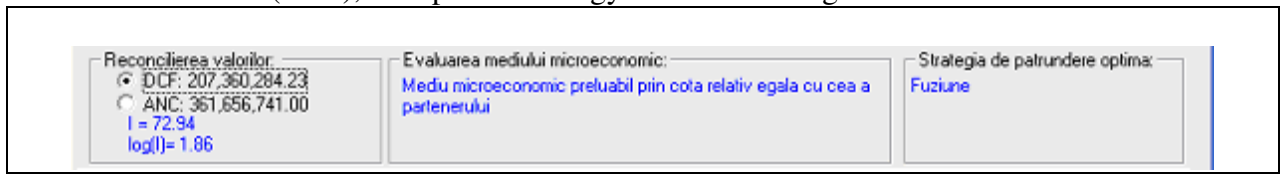

**Figure 4. The area of data input of comparative parameters (target market/country of origin) in AnBilanț software**

In our case, we can notice that for different values, the recommended strategy is different. For instance, in figure no. 5, we notice that, if we choose the value computed with Discounted Cash Flow method (DCF), the optimal strategy will be the merger.

**Figure 5. The area that displays the values calculated according to the Adjusted Net Asset method (ANC) in AnBilanț application**

In figure no. 6, one can notice that, if we choose the value calculated by using the Adjusted Net Asset (ANC) method the optimal entry strategy will be licensing, merger, strategic alliance or a management contract.

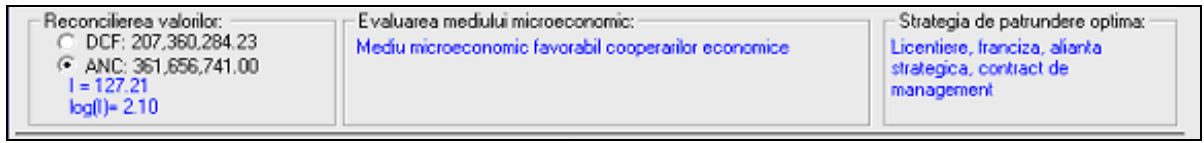

**Figure 6. The area that displays the values calculated according to the method Adjusted Net Asset method (ANC) in AnBilanț software**

Once fixed the rates and the parameters of the two regions and after determining the value that best describes the evolution of the enterprise, one can make the selection of the balance sheet elements and study their dynamics.

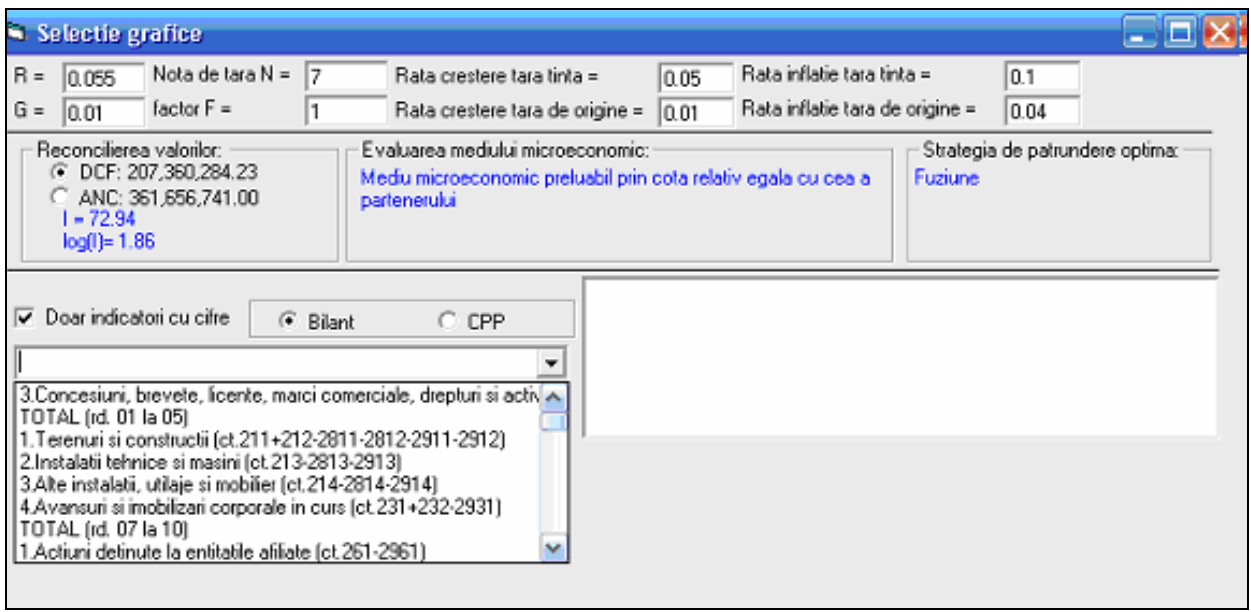

**Figure 7. The area of financial analysis applied for the selected patrimonial elements from the balance sheet or from the profit and loss account**

The selected item is transferred to the box on the right hand and remains on hold until the next action (acceptance and drawing a graph or de-select and delete from the box).

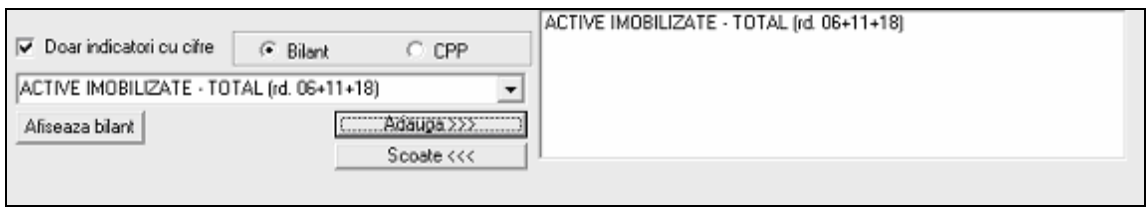

**Figure 8. Selection of an element from the balance sheet or from the profit and loss account**

In the case of accepting the selected items, by typing the button "Show balance sheet", the previously selected elements are shown graphically in their dynamics.

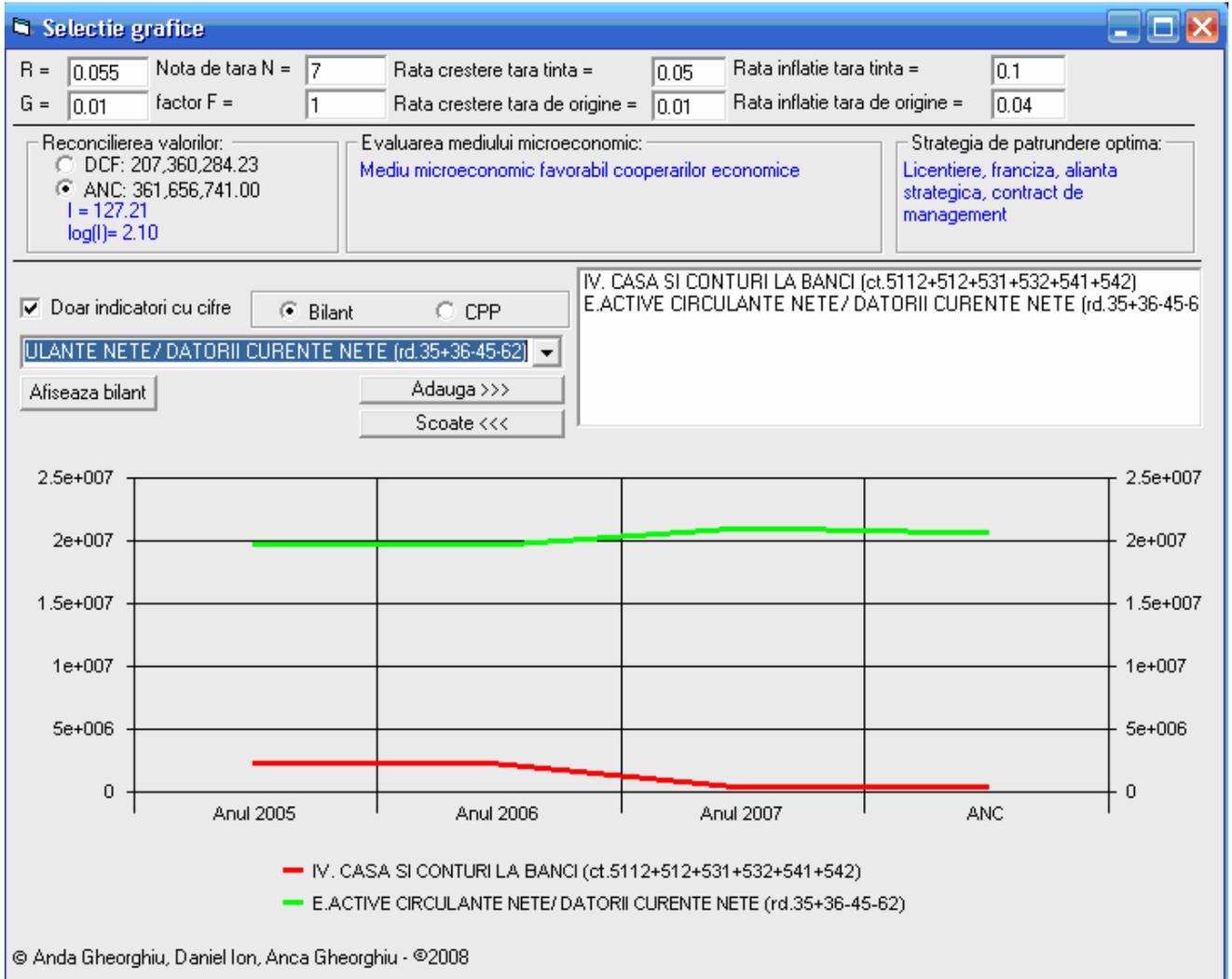

**Figure 9. The graphical dynamics of the balance sheet/P&L elements (I)**

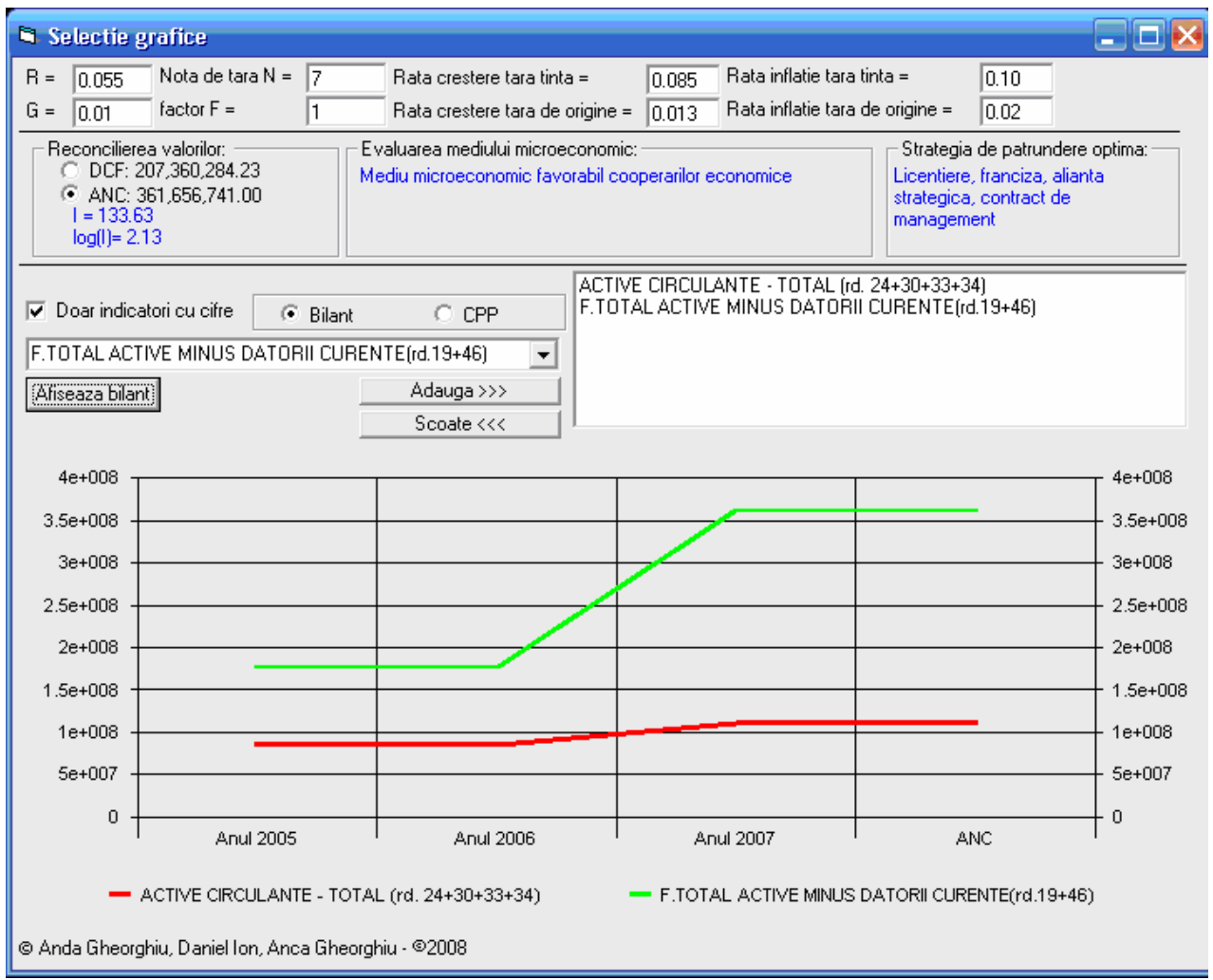

**Figure 10. The graphical dynamics of the balance sheet/P&L elements (II)**

## V. Conclusion

The risk management domain has increasingly become a key issue for any company; in this context, the complex case of market entry is still a puzzling issue for any manager who intends to expand the business internationally. The software application "AnBilanț" is meant to be an useful tool in order to take the right decision concerning the global market entry and to choose the optimal strategy, such as export, franchising, licensing, merger/acquisition or building a shining new industrial unit from scratch. AnBilanț is a valuable, user-friendly and reliable help in the decision process, especially in the case when the unit intended to be acquired is not listed at a stock market, and, consequently, it should be thoroughly evaluated by using one of the classical assessment methods like Discounted Cash Flow or Adjusted Net Asset method.

As AnBilanț is realized under Visual Basic 6 as an executable and runs under Windows operating system, it is able to identify and interrelate with various types of data files or Database Management Systems (i.e. software systems that allows users to save retrieve and modify information) such as Microsoft Access, dBase, FoxPro, Visual FoxPro, Paradox, SQL Server, etc; it allows the access to documents and Internet/Intranet applications.

# Bibliographic sources